\documentclass[aps,prd,showpacs,floatfix,superscriptaddress,nofootinbib,10pt,twocolumn]{revtex4}
\usepackage{color,amsmath,amssymb,graphicx,latexsym,subfigure}
\usepackage{threeparttable,multirow,txfonts,makecell,tabularx}
\DeclareMathAlphabet{\mathcal}{OMS}{cmsy}{m}{n}

\begin{document}
\title{Searching for cosmic string induced stochastic gravitational wave background with 
the Parkes Pulsar Timing Array}

\author{Ligong Bian}
\email{lgbycl@cqu.edu.cn}
\affiliation{Department of Physics and Chongqing Key Laboratory for Strongly Coupled Physics,
Chongqing University, Chongqing 401331, China}
\affiliation{Center for High Energy Physics, Peking University, Beijing 100871, China}

\author{Jing Shu}
\email{jshu@pku.edu.cn}
\affiliation{School of Physics and State Key Laboratory of Nuclear Physics and Technology, Peking University, Beijing 100871, China}
\affiliation{Center for High Energy Physics, Peking University, Beijing 100871, China}
\affiliation{CAS Key Laboratory of Theoretical Physics, Institute of Theoretical Physics, Chinese Academy of Sciences, Beijing 100190, China}
\affiliation{School of Physical Sciences, University of Chinese Academy of Sciences, Beijing 100049, China}
\affiliation{School of Fundamental Physics and Mathematical Sciences, Hangzhou Institute for Advanced\\ Study, University of Chinese Academy of Sciences, Hangzhou 310024, China}
\affiliation{International Centre for Theoretical Physics Asia-Pacific, University of Chinese Academy of Sciences, 100190 Beijing, China}

\author{Bo Wang}
\affiliation{International Centre for Theoretical Physics Asia-Pacific, University of Chinese Academy of Sciences, 100190 Beijing, China}

\author{Qiang Yuan}
\email{yuanq@pmo.ac.cn}
\affiliation{Key Laboratory of Dark Matter and Space Astronomy, Purple Mountain Observatory, Chinese Academy of Sciences, Nanjing 210023, China}
\affiliation{School of Astronomy and Space Science, University of Science and Technology of China, Hefei 230026, China}
    
\author{Junchao Zong}
\affiliation{Department of Physics, Nanjing University, Nanjing 210093, China}
\affiliation{CAS Key Laboratory of Theoretical Physics, Institute of Theoretical Physics, Chinese Academy of Sciences, Beijing 100190, China}

\begin{abstract}
We search for stochastic gravitational wave background emitted from cosmic strings using the
Parkes Pulsar Timing Array data over 15 years. While we find that the common power-law excess 
revealed by several pulsar timing array experiments might be accounted for by the gravitational 
wave background from cosmic strings, the lack of the characteristic Hellings-Downs correlation
cannot establish its physical origin yet. The constraints on the cosmic string model parameters 
are thus derived with conservative assumption that the common power-law excess is due to unknown
background. Two representative cosmic string models with different loop distribution functions
are considered. We obtain constraints on the dimensionless string tension parameter 
$G\mu<10^{-11}\sim10^{-10}$, which is more stringent by two orders of magnitude than that obtained 
by the high-frequency LIGO-Virgo experiment for one model, and less stringent for the other. 
The results provide the chance to test the Grand unified theories, with the spontaneous symmetry 
breaking scale of $U(1)$ being two-to-three orders of magnitude below $10^{16}$ GeV.
The pulsar timing array experiments are thus quite complementary to the LIGO-Virgo experiment 
in probing the cosmic strings and the underlying beyond standard model physics in the early Universe.   
\end{abstract}

\maketitle
\section{Introduction}

Cosmic strings (CS) are one-dimensional topological defects supposed to be 
formed during phase transitions where symmetry gets broken spontaneously \cite{Kibble:1976sj,Hindmarsh:1994re}.  
The CS dynamics can be described by the Nambu-Goto action for thin and local strings with no 
internal structures. In this situation, infinite strings can reach the scaling
regime \cite{Bennett:1987vf,Allen:1990tv,Sakellariadou:1990nd} and go to loops through the
intercommutation of intersecting string segments \cite{Vachaspati:1984dz}. The formed small 
loops will oscillate and emit gravitational wave (GW) bursts by the structures of Cusps and Kinks~\cite{Damour:2001bk,Damour:2000wa}. The superposition of uncorrelated 
GW bursts from many CSs form a stochastic gravitational wave background (SGWB).
In the Nambu-Goto string scenario, the SGWB is characterized by the loop number density and 
the string tension ($\mu$)~\cite{Vilenkin:2000jqa}. The dimensionless parameter $G\mu$ ($G$ is 
the Newtonian constant) that parameterized the gravitational interactions of strings is usually 
adopted in literature, which is tightly connect with the GUT scale ($M_{\rm GUT}\sim10^{16}$ GeV) 
as $G\mu\sim(\eta/M_{\rm GUT})^2$ since CSs are generally predicted in the symmetry breaking 
chain of grand unified theories (GUT). Therefore, the detection of SGWB from CSs provide an 
intriguing way to access the beyond-standard-model physics close to the GUT scale 
that are inaccessible by high-energy colliders \cite{King:2021gmj,King:2020hyd,Buchmuller:2019gfy,Caldwell:2022qsj}, such as leptogenesis 
and the type-I seesaw scale~\cite{Dror:2019syi}.    

The gravitational waves spectra from CSs span in a wide frequency range characterized by a plateau 
in the high-frequency region. The SGWB from CSs is one of the most promising target of LIGO-Virgo \cite{LIGOScientific:2021nrg,LIGOScientific:2017ikf}, Laser Interferometer Space Antenna (LISA) \cite{Auclair:2019wcv}, and pulsar timing arrays (PTA) \cite{Yonemaru:2020bmr,Sanidas:2012ee}. 
Recently, both LIGO-Virgo and PTA experiments made great progresses on the search of GWs. 
The LIGO-Virgo group place stringent constraints on the GW bursts and the SGWB from CSs in the 
high-frequency window \cite{LIGOScientific:2021nrg}. In the nanohertz range, several PTAs
reported the detection of a mysterious common red process 
\cite{NANOGrav:2020bcs,Goncharov:2021oub,Chen:2021rqp,Antoniadis:2022pcn},
which may be explained as SGWB from various sources including CSs  \cite{Ellis:2020ena,Bian:2020urb,Blasi:2020mfx,Blanco-Pillado:2021ygr}.
However, the GW interpretation of the common red process is still suspicious due to the
lack of the characteristic Hellings-Downs (HD) correlation in the data. 
In this Letter we search for SGWB signals generated from CSs utilizing the Parkes Pulsar 
Timing Array (PPTA) data. We consider two representative SGWB models of CSs with two different 
loop distribution functions, and derive the constraints on the dimensionless $G\mu$ parameter
of CS. The results can be translated to constraints on the spontaneous symmetry breaking scale 
of the GUT.

\section{SGWB spectra for cosmic string networks}

The SGWB from cosmic string networks comes from the uncorrelated superpositions of GW bursts of three contributions: cusps, kinks, and kink-kink collision. 
The SGWB spectrum emitted from cosmic strings is given by
\begin{equation}\label{eq:sgwb}
    \Omega_{\rm GW}(t_{0}, f) = \frac{f}{\rho_{c}}\ \frac{d\rho_{\rm GW}}{df}(t_{0}, f),
\end{equation}
where $\rho_{c}=\frac{3H_{0}^{2}}{8\pi G}$ is the critical energy density of the universe,
$\frac{d\rho_{\rm GW}}{df}(t_{0}, f)$ is the GWs energy density per unit frequency today. Considering different modes of loops oscillation ($n$), we have 
\begin{equation}\label{eq:dpf}
    \frac{d\rho_{\rm GW}}{df}(t_0,f)=G\mu^{2}\ \sum_{n}C_{n}(f)\ P_{n}\;,
\end{equation}
where the $C_n(f)$ is a function of loop distributions depending on the cosmological background. To account for all contributions from cusps, kinks, and kink-kink collisions, we adopt the $P_{n}$ from numerical simulations~\cite{Blanco-Pillado:2017oxo}. Analytically, $P_{n} = \frac{\Gamma}{\zeta(q)}n^{-q}$,   $\Gamma\approx 50$ and $\zeta(q)$ is the Riemann zeta function with $q = 4/3$, $5/3$, and $2$ 
representing for cusps, kinks, and kink-kink collisions, respectively. 

We first consider the SGWB from CSs with the loop production functions for non-self-intersecting 
loops being obtained directly from CS networks simulation in radiation and matter dominated 
era\footnote{It was noted that the inflation might also dilute the CS networks if CSs were 
formed before the onset of the inflation, in which case the nano-hertz GW experiments may not 
be sensitive to probe the SGWB from CSs \cite{Cui:2019kkd}.} by Blanco-Pillado, Olum, Shlaer \cite{Blanco-Pillado:2013qja,Blanco-Pillado:2011egf} (hereafter 
denoted as BOS model, which is dubbed as model {\tt A} by LIGO-Virgo group in Ref.~\cite{LIGOScientific:2021nrg}). In the model, one need to take into account contributions from the 
radiation dominated era and the matter dominated era with the loop number density $n(l,t)$ given 
in the {\tt Supplemental Material}. For the SGWB contributions from the radiation era, one has 
\begin{equation}
    C_{n}(f) = \frac{2n}{f^{2}}\int_{z_{\rm eq}}^{z_{\rm cut}}
    \frac{dz}{H_{0}\sqrt{\Omega_{r}}(1+z)^{8}}\ n_{r}(l,t),
\end{equation}
where $z_{\rm eq}$ is the redshift in the radiation-matter equality and $z_{\rm cut}$ is the cutoff redshift. The subscript $r$ here denotes radiation dominated era, similarly, in the following $rm$ is for the case where loops are formed in radiation dominated era but survive past to matter dominated era and $m$ for the case of matter dominated era. We note that here the $n_{r}(l,t)$ is the loop distribution function rather than the oscillation mode $n$ before the integration.  
The SGWB contributions from matter dominated era constitute two parts: loops that survive to matter dominated era and loops formed in matter dominated era. The $C_{n}$ takes the form of 
\begin{equation}
    C_{n}(f) = \frac{2n}{f^{2}}\int_{0}^{z_{\rm eq}}
    \frac{dz}{{H_{0}\sqrt{\Omega_{m}}}(1+z)^{15/2}}\ n_{i}(l,t),
\end{equation} 
with $i$ being $rm$ and $m$ for these two cases, respectively.

Following the same procedure, we can calculate the SGWB spectrum with loop distribution functions 
being derived analytically by Lorenz, Ringeval, and Sakellariadou~\cite{Lorenz:2010sm} (denoted
as LRS, which is dubbed as model {\tt B} by LIGO-Virgo group in Ref.~\cite{LIGOScientific:2021nrg}), where the distribution of non-self-intersecting scaling loops are extracted from
simulations~\cite{Ringeval:2005kr}. In the model, the contribution from the radiation dominated 
era takes the form of
\begin{equation}
    \Omega_{\rm GW}(f)=\frac{64\pi G^2\mu^2\Omega_r}{3}\sum_n P_n \int dx\ n(x),
\end{equation}
and in the matter dominated era, the GW spectrum is
\begin{equation}
    \Omega_{\rm GW}(f)=\frac{162\pi G^2\mu^2}{\Omega_m^{-2}H_0^{-2}}\frac{1}{f^2}\sum_n\ P_n\ n^2\int_{}^{}dx\ n(x).
\end{equation}

\begin{figure}[!htp]
    \centering
    \includegraphics[width=8.5cm]{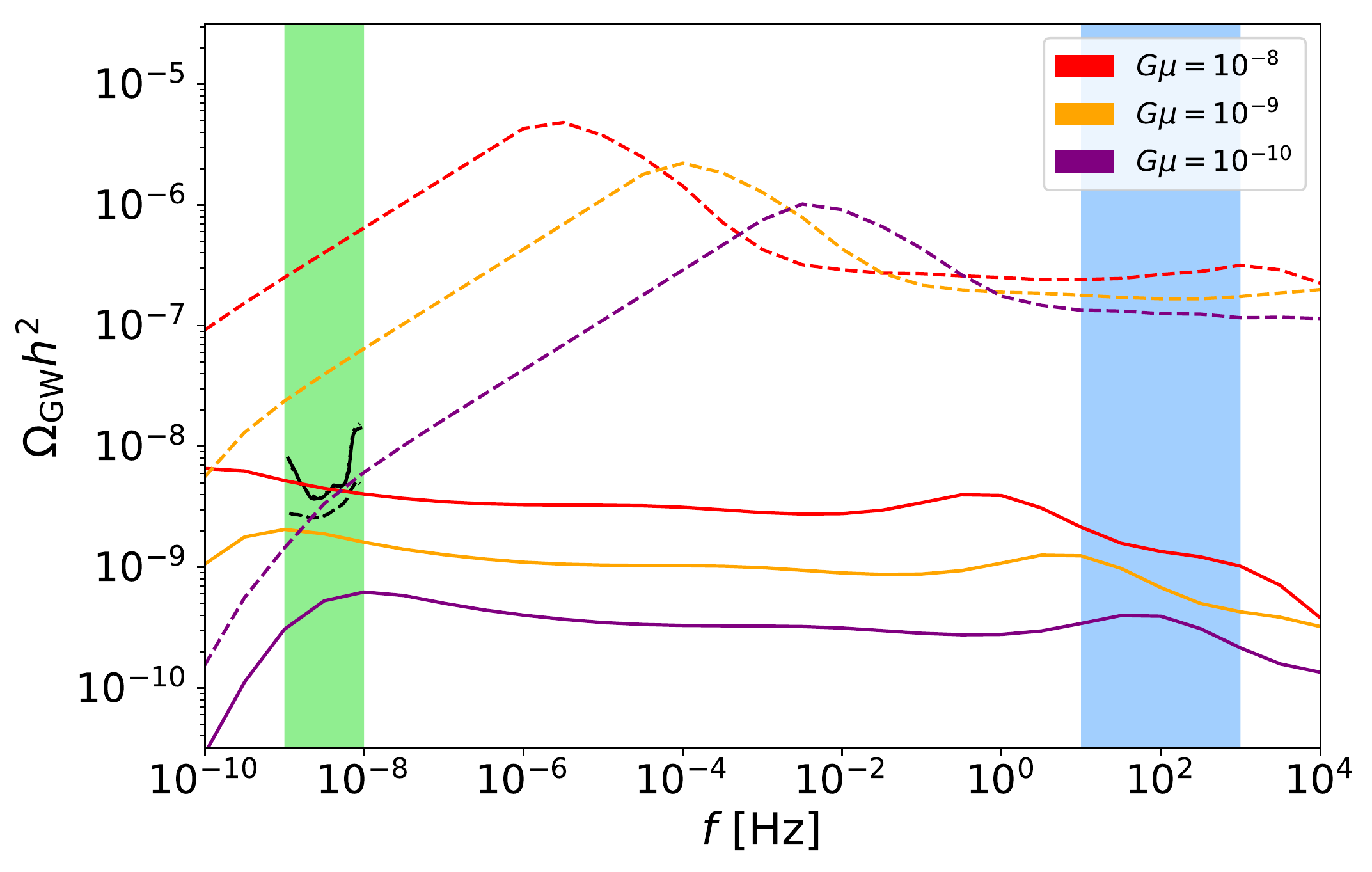}
    \caption{Expected spectra of SGWB from CS for the BOS (LRS) model with $G\mu=10^{-8}$, 
    $10^{-9}$, and $10^{-10}$, which are indicated by red, orange and purple solid (dashed) lines. 
    The solid and dashed black lines represent the $95\%$ C.L. upper limits for free 
    spectrum assumption derived from the PPTA data for different hypotheses (H2/H3 and H4 
    in Tables \ref{BOSBYS} and \ref{LRSBYS}, respectively). The sensitive frequency regions 
    of PTA and LIGO-Virgo are denoted by green and blue shaded bands.}\label{FSGW}
\end{figure}

Fig.~\ref{FSGW} shows the total SGWB spectra of the CS-induced SGWB for the BOS and LRS models,
for a few values of $G\mu$. It is shown that the BOS model predicts relatively flat spectra 
(solid lines) and hence the PTA experiment is more suitable to probe it. On the other hand,
the LRS model predicts harder spectra (dashed lines) which is more optimal for the high-frequency
GW experiment such as LIGO-Virgo. Also shown in Fig.~\ref{FSGW} are the upper limits of the 
free-spectrum SGWB derived from the PPTA data \cite{Xue:2021gyq}. We find that the CS models 
with $G\mu$ of $10^{-8}\sim10^{-10}$ are severely constrained by the PPTA data. 
The magnitude of the SGWB spectra for the LRS model is much higher than that of the BOS model 
at high frequencies due to more small loops in the loop distribution function dominate in such 
frequency range.


\section{Data Analysis}

The dataset we used in this work is the second data release (DR2) of the PPTA \cite{Kerr:2020qdo}, 
which is available in the CSIRO pulsar data archive\footnote{https://doi.org/10.4225/08/5afff8174e9b3}. 
Searches for SGWBs and other fundamental physics problems using these data (or the subsets) were
also carried out \cite{Shannon:2015ect,Porayko:2018sfa,PPTA:2021uzb,Goncharov:2021oub,Xue:2021gyq}.
In PTA, to search for such an SGWB signal from CS is to find spatially correlated time residuals
among time-of-arrivals (ToA) of different pulsars. The residuals are composed of deterministic 
timing models, noises, and the hypothetical signal (SGWB from CS here). 
We use the {\tt{TEMPO2}} tool \cite{Hobbs:2006cd,Edwards:2006zg} to fit the timing models of
pulsars, the {\tt{ENTERPRISE}}\footnote{https://github.com/nanograv/enterprise} and the
{\tt{ENTERPRISE\_EXTENSIONS}}\footnote{https://github.com/nanograv/enterprise\_extensions} 
to model the noises, and {\tt{PTMCMCSampler}} \cite{justin_ellis_2017_1037579} to do the 
Bayesian analysis. The noise model is based on single pulsar analyses of Ref.~\cite{Goncharov:2020krd}.
Briefly speaking, the stochastic noises mainly include two parts, the white noise and red noise. 
The white noise may come from the radio frequency interference, pulse profile changes or 
instrumental artifacts. We use three parameters, {\tt EFAC} (Error FACtor), {\tt EQUAD}
(Error added in QUADrature) and {\tt ECORR} (Error of CORRelation between ToAs in a single epoch), to account for white noise which can not be subtracted by fitting 
the timing model. As usually done \cite{Porayko:2018sfa,PPTA:2021uzb}, we fix the white noise 
parameters as their maximum likelihood values from the single pulsar analyses in the Bayesian
analysis. The red noises are mainly caused by the irregularities of the pulsar spin (spin noise) 
and the dispersion measure of photons when traveling through the interstellar medium (DM noise).
For some pulsars there are band noise for ToAs of a certain photon frequency band and
chromatic noise which correlates between different photon frequencies. All the red noises
are modeled as power-law forms with amplitude $A$ and slope $\gamma$. 
Several analyses revealed that there is a common power-law (CPL) process in the pulsar ToAs,
whose nature is still in debate \cite{NANOGrav:2020bcs,Goncharov:2021oub,Chen:2021rqp,Antoniadis:2022pcn}. 
In this work we will test different assumptions on the CPL, as either the SGWB signal from
CS (see also \cite{Ellis:2020ena,Bian:2020urb}) or an unknown background.
The solar system ephemeris uncertainties are modeled with a 11-parameter model
{\tt BayesEphem} implemented in {\tt{ENTERPRISE}}, including perturbations in masses of
major planets, the drift rate of the Earth-Moon barycenter orbit, and the perturbations
of the Earth's orbit from Jupiter's average orbital elements described by 6 parameters 
\cite{NANOGRAV:2018hou}.
We summarize the noise and signal parameters together with their priors adopted in the 
Bayesian analysis in Table~\ref{PPT} in the {\tt Supplemental Material}.

\section{RESULT}


To address how significant the CS-induced SGWB signals in the data, we test the Bayes factor (BF)
of the ``signal hypothesis'' against the null hypothesis, following the Savage-Dickey 
formula \cite{10.2307/2958475}
\begin{equation}
{\rm BF}_{10} = \frac{P_{1}(\boldsymbol{D})}{P_{0}(\boldsymbol{D})} = \frac{P(\boldsymbol{\phi}=\boldsymbol{\phi_0})}{P(\boldsymbol{\phi}=\boldsymbol{\phi_0}|\boldsymbol{D})} ,
\end{equation}
where ${\rm BF}_{10}$ means the BF of hypothesis H1 against H0, $\boldsymbol{D}$ is the 
observational data, $\boldsymbol{\phi}$ is the parameters of the signal model, and 
$\boldsymbol{\phi_0}$ is the parameters of the null hypothesis which is a subset of 
$\boldsymbol{\phi}$. $P_{0}$ and $P_{1}$ are the evidence of the noise and signal hypotheses. 
The ${\rm BF}_{10}$ is equivalent to the ratio of the prior to the posterior probabilities
of the null hypothesis. 

The null hypothesis (H0) corresponds to the model with only the pulsar timing model and noise.
For hypothesis H1, we assume an additional CPL process in the model. The CS model with the 
HD correlation is included in substitution of the CPL process is labelled as H2.
In addition, in H3 and H4 we consider simultaneously the CPL process as an unknown systematics
and the CS contribution, in which the auto-correlation of a pulsar's own ToAs is not subtracted
(H3) and subtracted (H4), respectively. The results of the fittings for different hypotheses
are summarized in Tables \ref{BOSBYS} and \ref{LRSBYS}.

\begin{table*}[!htp]
    \caption{Hypotheses, Bayes factors, and estimated model parameters for the BOS model.}
    \label{BOSBYS}
    \begin{tabular}{|l|c|c|c|c|c|c|}
    \hline 
    \multirow{2}{*}{Hypothesis} & 
    Pulsar &
    CPL &
    HD process  &
    \multirow{2}{*}{Bayes Factors}  & \multicolumn{2}{c|}{Parameter Estimation ($1\sigma$ interval)}\\
    \cline{6-7}
    & Noise & Process &CS spectrum & & $\quad\quad \log_{10}G\mu \quad\quad$ & $\log_{10}A_{\text{CPL}}, \gamma_{\text{CPL}}$\\
    \hline
    H0:Pulsar Noise & \checkmark & & & & &  \\
    \hline
    H1:CPL & \checkmark & \checkmark & & $10^{3.2}$ (/H0) & & $-14.48^{+0.62}_{-0.64}, 3.34^{+1.37}_{-1.53}$ \\
    \hline
    H2:CS & \checkmark &  & \checkmark(full HD) & $10^{3.1}$ (/H0) & $-10.38^{+0.21}_{-0.21}$ & \\
    \hline
    H3:CS1 & \checkmark & \checkmark & \checkmark(full HD) & $1.96\ $ (/H1) & $<-10.02$ (95\% C.L.) & $-15.58^{+1.21}_{-1.64}, 3.11^{+1.95}_{-2.02}$ \\
    \hline
    H4:CS2 & \checkmark & \checkmark & \checkmark(no-auto HD) & $0.60\ $ (/H1) & $< -10.54$ (95\% C.L.) & $-14.61^{+0.58}_{-0.59}, 3.63^{+1.24}_{-1.40}$ \\
    \hline
    \end{tabular}
\end{table*}

\begin{table*}[!htp]
    \caption{Hypotheses, Bayes factors, and estimated model parameters for the LRS model.}
    \label{LRSBYS}
    \begin{tabular}{|l|c|c|c|c|c|c|}
    \hline 
    \multirow{2}{*}{Hypothesis} & 
    Pulsar &
    CPL &
    HD process  &
    \multirow{2}{*}{Bayes Factors}  & \multicolumn{2}{c|}{Parameter Estimation ($1\sigma$ interval)}\\
    \cline{6-7}
    & Noise & process  &CS spectrum & & $\quad\quad \log_{10}G\mu \quad\quad$ & $\log_{10}A_{\text{CPL}}, \gamma_{\text{CPL}}$\\
    \hline
    H0:Pulsar Noise & \checkmark & & & & &  \\
    \hline
    H1:CPL & \checkmark & \checkmark & & $10^{3.2}$ (/H0) & & $-14.48^{+0.62}_{-0.64}, 3.34^{+1.37}_{-1.53}$ \\
    \hline
    H2:CS & \checkmark &  & \checkmark(full HD) & $10^{3.3}$ (/H0) & $-10.89^{+0.14}_{-0.17}$ & \\
    \hline
    H3:CS1 & \checkmark & \checkmark & \checkmark(full HD) & $ 1.62 \ $ (/H1) & $<-10.64$ (95\% C.L.) & $-15.44^{+1.18}_{-1.74}, 3.08^{+1.94}_{-1.99}$ \\
    \hline
    H4:CS2 & \checkmark & \checkmark & \checkmark(no-auto HD) & $0.55 \ $ (/H1) & $<-11.04$ (95\% C.L.) & $-14.57^{+0.58}_{-0.59}, 3.54^{+1.24}_{-1.41}$ \\
    \hline
    \end{tabular}
\end{table*}

\begin{figure}[!htp]
    \centering
    \includegraphics[width=8.5cm]{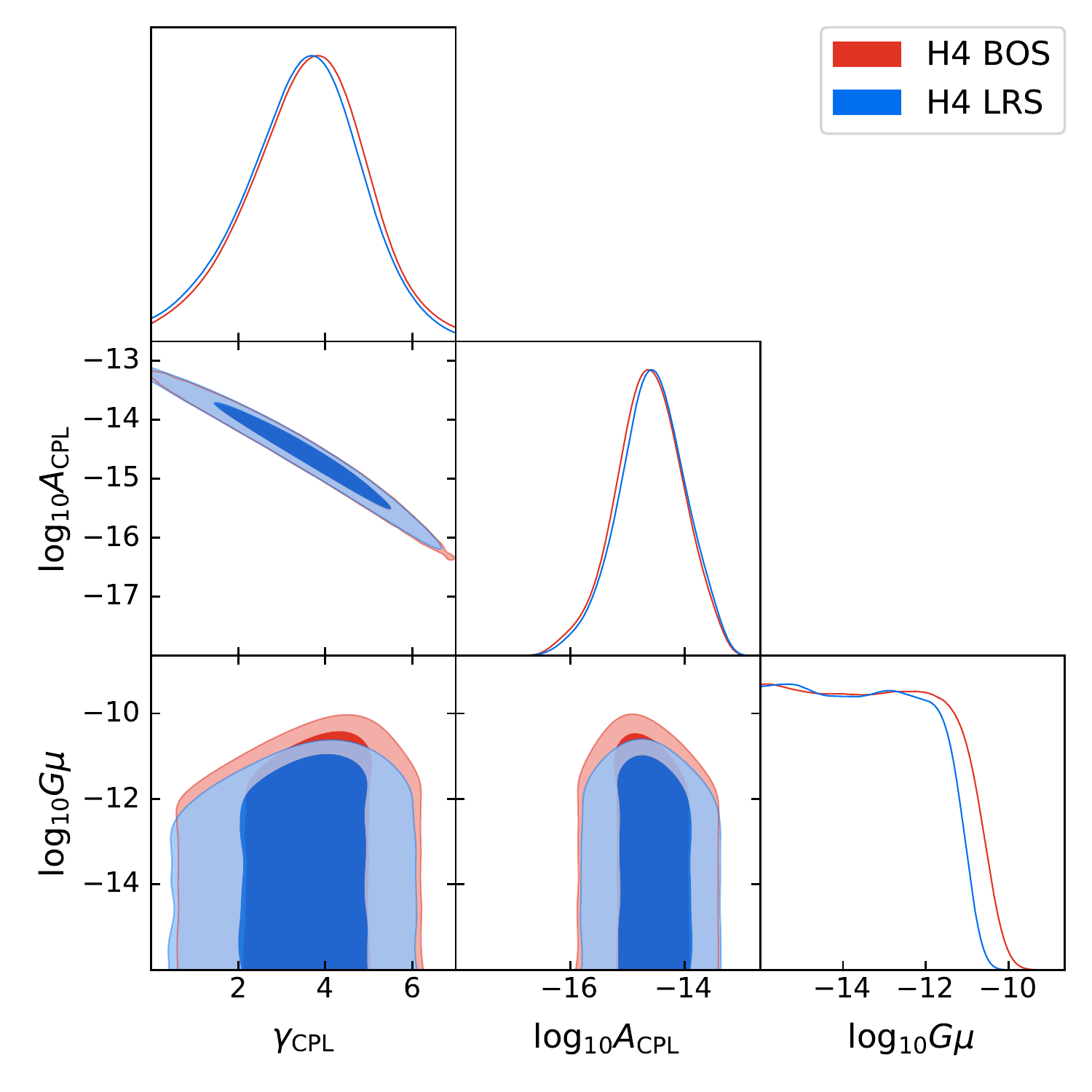}
    \caption{Posteriors distribution of $\log_{10}G\mu$ and the CPL parameters of BOS 
    (red) and LRS (blue) models assuming a no-auto HD correlation. The $1\sigma$ and 
    $2\sigma$ regions are presented in light and dark colors.}
    \label{H4BL}
\end{figure}

For hypothesis H1, a clear CPL signal with a BF of $10^{3.2}$ is revealed in the data,
consistent with previous studies \cite{NANOGrav:2020bcs,Goncharov:2021oub,Chen:2021rqp,Antoniadis:2022pcn}. 
The posterior distributions of the CPL model parameters are shown in Fig.~\ref{H1}
in the {\tt Supplemental Material}. Similar signal with comparable BF value is found
if we assume a CS-induced SGWB component in the model rather than the CPL (H2).
Fitting results of the CS parameter $\log_{10} G\mu$ are $-10.38\pm0.21$ and 
$-10.89^{+0.14}_{-0.17}$ for the BOS and LRS model, with $1\sigma$ error bars presented. 
The corresponding parameter values are consistent with that employed to account for the
NANOGrav data, but with different model assumptions \cite{Ellis:2020ena}.
Since the nature of the CPL process is unclear, and particularly the characteristic HD
correlation of GWs is still lack, we also test the hypotheses that treating the CPL as 
an unknown background. The resulting BF of hypotheses H3 or H4 against H1 is close to 1
which means that the evidence of the CS is not significant in case of a CPL background.
We therefore derive the 95\% credible level (C.L.) upper limits on $\log_{10}G\mu$, 
as given in Tables \ref{BOSBYS} and \ref{LRSBYS}. The posterior distributions of the CPL 
and CS parameters for H4 are shown in Fig.~\ref{H4BL}. 
More results of the analysis based on different hypotheses can be found in 
Figs.~\ref{CSH2}-\ref{H3BL} in the {\tt Supplemental Material}.
Additional tests assuming that the CPL has an astrophysical origin from the supermassive
binary black hole (SMBBH) give similar conclusion (see the {\tt Supplemental Material}).

\begin{figure}[!htp]
    \centering
    \includegraphics[width=8.5cm]{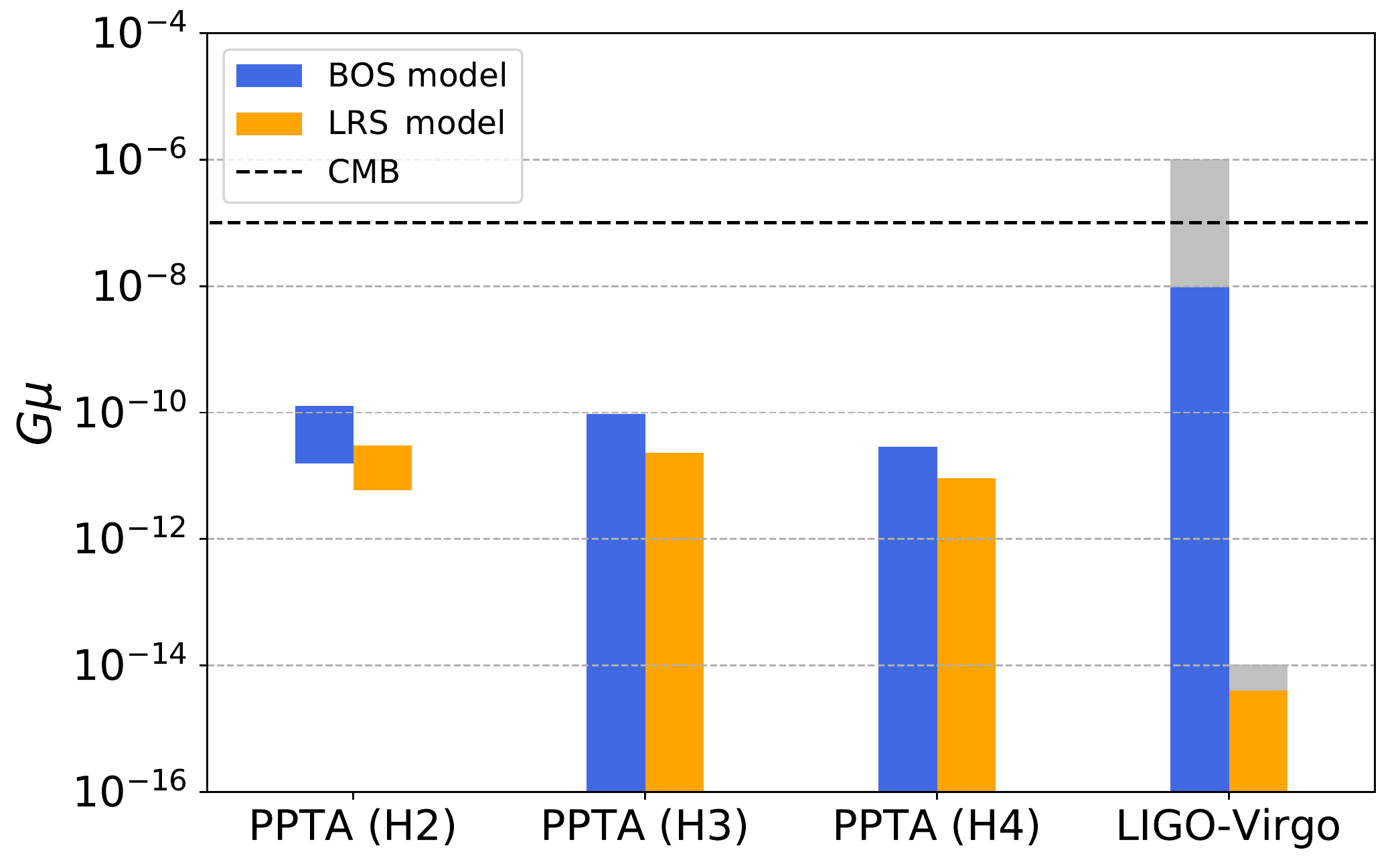}
    \caption{Constraints on $G\mu$ at the $95\%$ C.L. from the PPTA data for different hypotheses, 
    compared with those obtained by LIGO-Virgo experiment~\cite{LIGOScientific:2021nrg}. 
    Silver boxes represent the ranges reported in the LIGO-Virgo analysis. The limit from cosmic microwave background (CMB) is also reported by the black dashed line~\cite{Planck:2013mgr}.}
    \label{COMP}
\end{figure}

Our results can be compared with those obtained by the LIGO-Virgo observations of high-frequency
GWs \cite{LIGOScientific:2021nrg}, which is displayed in Fig.~\ref{COMP}. We also show the  
bound from the cosmic microwave background (CMB), while the less stronger limit from Big Bang nucleosynthese (BBN) is too weak to be shown in the 
figure~\cite{Planck:2013mgr,LIGOScientific:2021nrg}. 
The LIGO-Virgo upper limits of $G\mu$ are $10^{-8}-10^{-6}$ and $(4.0-6.3)\times10^{-15}$ 
for the BOS and LRS model (model A and B in \cite{LIGOScientific:2021nrg}), respectively. 
Compared with LIGO-Virgo results, the PPTA upper limits on $\log_{10}G\mu$ improve by more 
than two orders of magnitude for the BOS model. For the LRS model, the LIGO-Virgo result is 
more stringent because the SGWB spectrum is harder. The PTA experiment is thus very 
complementary to the LIGO-Virgo experiment for the searches of CS-induced SGWB.

\section{CONCLUSION}
The very precise timing measurements of pulsars provide a powerful tool to probe the 
fundamental new physics process occurred in the early Universe. In this work we use
more than 15 years of observations of 26 millisecond pulsars by the PPTA experiment to
search for SGWB from CS networks. We show that the CPL process revealed by several PTAs recently 
can be explained by the CS-induced SGWB signal with $\log_{10}G\mu\sim -10.38$ and $-10.89$ 
for the two typical classes of CS models, the BOS and LRS models. While the LRS model 
explanation would be excluded by the LIGO-Virgo observations at high frequencies, the 
BOS model explanation is consistent with the LIGO-Virgo data \cite{LIGOScientific:2021nrg}.
As a conservative alternative, we also assume the CPL as a background component and 
turn to set limits on the CS model parameters. The PPTA upper limits on $\log_{10}G\mu$
reach about $-10\sim-11$, depending on the CS models and the analysis methods. 
These constraints correspond to the symmetry breaking scale $\eta\sim 10^{13}-10^{14}$ GeV, 
that challenge the grand unification theories below the GUT scale already.
For the BOS model, the PPTA limits are more stringent than those from LIGO-Virgo,
and for the LRS model the LIGO-Virgo experiment is more sensitive. This is due to the fact that, 
in the LRS model, more small loops are contained in the loop distribution functions that control 
the behaviours of the SGWB spectra from CS networks in high frequency range. We expect that
continuous accumulation of more precise data by PTAs around the world in the near
future will critically test the CS models and more generally, a class of GUTs admitting 
the U(1) symmetry breaking in the early Universe.


\section{Acknowledgements}
This work uses the public data from the Parkes Pulsar Timing Array. We thank Huai-Ke Guo and Yue Zhao for useful discussions, and appreciate Xiao Xue for the help on the data analysis in the early stage of this work. L.B. is supported by the National Key Research and Development Program of China under Grant No. 2021YFC2203004, the National Natural Science Foundation of China (NSFC) under Grants No. 12075041 and No. 12047564, the Fundamental Research Funds for the Central Universities of China under Grants No. 2021CDJQY-011 and No. 2020CDJQY-Z003, and Chongqing Natural Science Foundation under Grant No. cstc2020jcyj-msxmX0814. J. S. is supported by the NSFC under Grants No. 12025507, No. 11690022, and No. 11947302, by the Strategic Priority Research Program and Key Research Program of Frontier Science of the Chinese Academy of Sciences (CAS) under Grants No. XDB21010200, No. XDB23010000, No. XDPB15, and No. ZDBS-LY-7003, and by the CAS Project for Young Scientists in Basic Research under Grant No. YSBR-006. Q.Y. is supported by the Program for Innovative Talents and Entrepreneur in Jiangsu and the Key Research Program of CAS under Grant No. XDPB15.

\bibliography{refs}
\bibliographystyle{ieeetr}

\clearpage

\setcounter{figure}{0}
\renewcommand\thefigure{S\arabic{figure}}
\setcounter{table}{0}
\renewcommand\thetable{S\arabic{table}}
\setcounter{equation}{0}
\renewcommand\theequation{S\arabic{equation}}

\section*{Supplemental material}

\subsection{CS loop distributions}

{\it The BOS model.} --- In this model, the loop production functions are inferred from Nambu-Goto simulations of CS networks in radiation and matter dominated era directly. 
In the radiation dominated era, we take $l/t \leqslant 0.1$ to consider the cutoff of the maximum size of loops. In the matter dominated era, as suggested by simulations, 
plenty of loops formed in the radiation era will survive till the time of 
radiation-matter equality and emit GWs in the matter dominated era, which imposes a constrain on loop size, $l/t < 0.09 t_{\rm eq}/t - \Gamma G\mu$. 
Furthermore, loops can also be produced when the CS networks reach the scaling regime in the matter dominated era, in which case the loop size should satisfy $l/t < 0.18$. The loop distribution functions of all the circumstances are 
\begin{eqnarray}
 n_{r}(l, t)&=&\frac{0.18}{t^{3/2}(l + \Gamma G\mu t)^{5/2}},  l/t \leqslant 0.1,\\
 n_{rm}(l, t)&=&\frac{0.18t_{\rm eq}^{1/2}}{t^2\ (l + \Gamma G\mu t)^{5/2}},\, l/t < 0.09 t_{\rm eq}/t - \Gamma G\mu,\\
 n_{m}(l, t)&=&\frac{0.27-0.45(l/t)^{0.31}}{t^{2}(l + \Gamma G\mu t)^{2}}, l/t < 0.18\;.
\end{eqnarray}

{\it The LRS model.} --- Different from the BOS model, in the LRS model, the loop distribution 
of the non-self-intersecting scaling loops rather than the loop production function is inferred 
from CS network simulations \cite{Lorenz:2010sm}. On scales $x = l/t \gg \Gamma G\mu \equiv x_d$, 
the simulation gives 
\begin{equation}\label{nx0}
    n(x) = \frac{C_0}{x^p},
\end{equation}
where the two constants $C_0$ and $p$ in the radiation dominated era and matter dominated era are 
\begin{align}
    p = 0.60^{+0.21}_{-0.15}\big{|}_r,&\quad p = 0.41^{+0.08}_{-0.07}\big{|}_m,&\\
    C_0 = 0.21^{-0.12}_{+0.13}\big{|}_r,&\quad C_0 = 0.09^{-0.03}_{+0.03}\big{|}_m.&
\end{align}
Note that, not only loops would emit GWs — which decreases their length $l$ — but the GW 
emission also back-reacts on the loops. The back-reaction smooths out the loops on the smallest 
scales (in particular the kinks), hindering the formation of smaller loops. To extend the loop
distribution given above down to smaller scales, a further length scale $x_c \ll x_d$ 
(the so-called ``gravitational back-reaction scale'') is introduced, which is estimated as 
$x_c = 20(G\mu)^{1+2\chi}$ through matching the loop distribution on scales $x\gg x_d$ 
(Eq.~(\ref{nx0})), where $\chi_r=0.2^{+0.07}_{-0.10}$ for the radiation dominated era and
$\chi_m=0.295^{+0.03}_{-0.04}$ for the matter dominated era \cite{Auclair:2019wcv}.
The specific distributions in all periods are then given by~\cite{Lorenz:2010sm}
\begin{eqnarray}
n(x > x_d) & \simeq & \frac{C}{(x+x_d)^{3-2\chi}},\\
n(x_c< x < x_d) & \simeq & \frac{C(3\nu-2\chi-1)}{2-2\chi}\frac{1}{x_d}\frac{1}{x^{2(1-\chi)}},\\
n(x < x_c < x_d) & \simeq & \frac{C(3\nu-2\chi-1)}{2-2\chi}\frac{1}{x_c^{2(1-\chi)}}\frac{1}{x_d},
\end{eqnarray}
where $C=C_0(1-\nu)^{2-p}$, and $\nu=1/2$ for radiation dominated era and $\nu=2/3$ for
matter dominated era.

\subsection{PTA model parameters}
Table \ref{PPT} summarizes the major model parameters and their prior distributions of the 
noise and signal components used in the analysis. The timing model parameters of each pulsar
are not listed here.

\begin{table*}[htp!]
\caption{Parameters and their prior distribution in data analysis. U and log-U stand for the uniform and log-uniform distribution.}
\label{PPT}
\centering\setlength{\tabcolsep}{2.5mm}

\begin{tabular}{lccc}
\hline\hline
parameter            & \multicolumn{1}{c}{Description}                      & \multicolumn{1}{c}{Prior}                         & \multicolumn{1}{c}{Comments}                 \\ \hline
\multicolumn{4}{c}{Noise parameters($\boldsymbol{\vartheta}$)}                                                                                                                                       \\
EFAC                 & \multicolumn{1}{c}{White-noise modifier per backend} & \multicolumn{1}{c}{U $[0, 10]$}                 & \multicolumn{1}{c}{Fixed for setting limits} \\
EQUAD                & \multicolumn{1}{c}{Quadratic white noise per backend} & \multicolumn{1}{c}{log-U $[-10, -5]$}           & \multicolumn{1}{c}{Fixed for setting limits} \\
ECORR                & \multicolumn{1}{c}{Correlated-ToAs white noise per backend} & \multicolumn{1}{c}{log-U $[-10, -5]$}           & \multicolumn{1}{c}{Fixed for setting limits} \\

$A_{\textrm{SN}}$             & \multicolumn{1}{c}{Spin-noise amplitude}             & \multicolumn{1}{c}{log-U $[-20, -6]$ (search)} & \multicolumn{1}{c}{One parameter per pulsar} \\
                     & \multicolumn{1}{c}{}                                 & \multicolumn{1}{c}{U [$10^{-20}, 10^{-6}$] (limit)}  & \multicolumn{1}{c}{}                         \\
$\gamma_{\textrm{SN}}$        & Spin-noise spectral index                            & U $[0, 10]$                                      & One parameter per pulsar                     \\
$A_{\textrm{DM}}$             & DM-noise amplitude                                   & log-U $[-20, -6]$ (search)                    & One parameter per pulsar                     \\
                     &                                                      & \multicolumn{1}{c}{U [$10^{-20}, 10^{-6}$] (limit)}                                      &                                              \\
$\gamma_{\textrm{DM}}$        & DM-noise spectral index                            & U $[0, 10]$                                      & One parameter per pulsar                     \\

$A_{\textrm{BAND}}$           & Band-noise amplitude 
                  & log-U $[-20, -6]$ (search)                    & One parameter partial pulsars                     \\
                     &                                                      & \multicolumn{1}{c}{U [$10^{-20}, 10^{-6}$] (limit)}                                      &                                              \\
$\gamma_{\textrm{BAND}}$           & Band-noise spectral index
                  &  U $[0, 10]$                    & One parameter  partial pulsars                     \\                
          & One parameter per pulsar                     \\
$A_{\textrm{CHROM}}$             & Chromatic-noise amplitude                                   & log-U $[-20, -6]$ (search)                    & One parameter partial pulsars                     \\
                     &                                                      & \multicolumn{1}{c}{U [$10^{-20}, 10^{-6}$] (limit)}                                      &                                              \\
$\gamma_{\textrm{CHROM}}$        & Chromatic-noise spectral index                            & U $[0, 10]$                        & One parameter partial pulsars                     \\
$n_{\textrm{CHROM}}$        & Index of chromatic effects                            & U $[0, 6]$                        & Fixed for single pulsar                     \\
$A_{\textrm{CPL}}$             & CPL process amplitude                                   & log-U $[-18, -11]$ (search)                    & One parameter per PTA                     \\
                     &                                                      & \multicolumn{1}{c}{U [$10^{-18}, 10^{-11}$] (limit)}                                      &                                              \\
$\gamma_{\textrm{CPL}}$             & CPL process power index                                   & U $[0, 7]$                     & One parameter per PTA                     \\

\hline                    
\multicolumn{4}{c}{Signal parameters ($\boldsymbol{\psi}$)}                                                                                                                                      \\
$G\mu$  & String tension                                       & log-U $[-16, -7]$ (search)                      & One parameter per PTA                        \\
                     &                                                      & U [$10^{-16}, 10^{-7}$] (limit)               &                                              \\
\hline
\multicolumn{4}{c}{BayesEphem parameters ($\boldsymbol{\phi}$)}                                                                                                                               \\
$z_{\rm drift}$          & Drift-rate of Earth’s orbit about ecliptic $z$-axis & U$[-10^{-9}, 10^{-9}]\ {\rm rad}\ {\rm yr}^{-1}$             & One parameter per PTA                        \\
$\Delta M_{\rm Jupiter}$ & Perturbation of Jupiter's mass                       & $N(0,1.5\times 10^{-11})~M_{\bigodot}$          & One parameter per PTA                        \\
$\Delta M_{\rm Saturn}$  & Perturbation of Saturn's mass                        & $N(0, 8.2\times 10^{-12})~M_{\bigodot}$          & One parameter per PTA                        \\
$\Delta M_{\rm Uranus}$  & Perturbation of Uranus' mass                         & $N(0, 5.7\times 10^{-11})~M_{\bigodot}$          & One parameter per PTA                        \\
$\Delta M_{\rm Neptune}$ & Perturbation of Neptune's mass                       & $N(0, 7.9\times 10^{-11})~M_{\bigodot}$          & One parameter per PTA                        \\
$PCA_i$                  & Principal components of Jupiter's orbit              & U $[-0.05, 0.05]$                                          & One parameter per PTA                        \\
\hline\hline
\end{tabular}
\end{table*}

\subsection{Results of other Hypotheses}

\begin{figure}[!htp]
\includegraphics[width=0.48\textwidth]{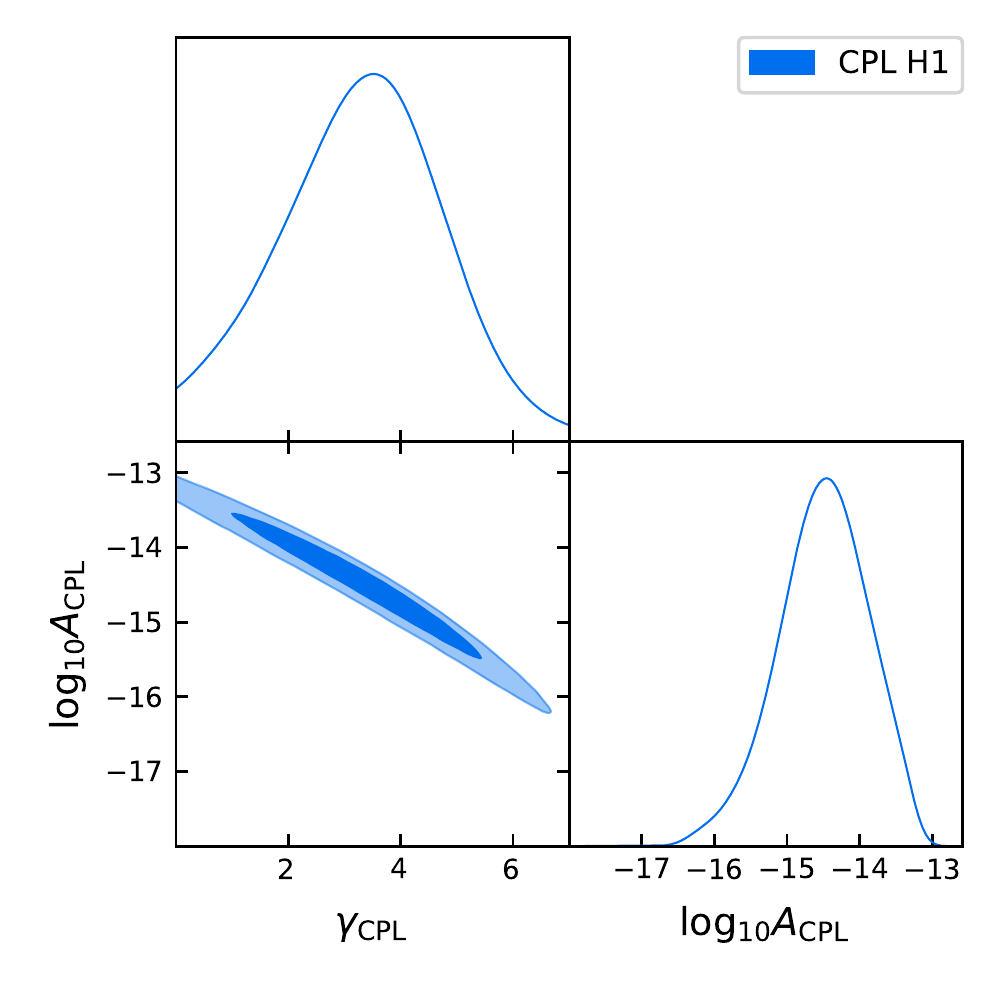}
\caption{Posterior distributions of the amplitude $\log_{10}A_{\textrm{CPL}}$ and power-law index $\gamma_{\textrm{CPL}}$ of the CPL component for hypothesis H1.}
\label{H1}
\end{figure}

Fig.~\ref{H1} shows the parameter distributions of the CPL process (H1). The best-fitting
values and $1\sigma$ errors of the amplitude and spectral index are $\log_{10}A_{\textrm{CPL}} 
= -14.48^{+0.62}_{-0.64}$ and $\gamma_{\textrm{CPL}} = 3.34^{+1.37}_{-1.53}$. The BF of the
hypothesis H1 against the null hypothesis H0 is $10^{3.2}$, indicating a strong evidence of 
the exist of such a component.

\begin{figure}[!htp]
\includegraphics[width=0.45\textwidth]{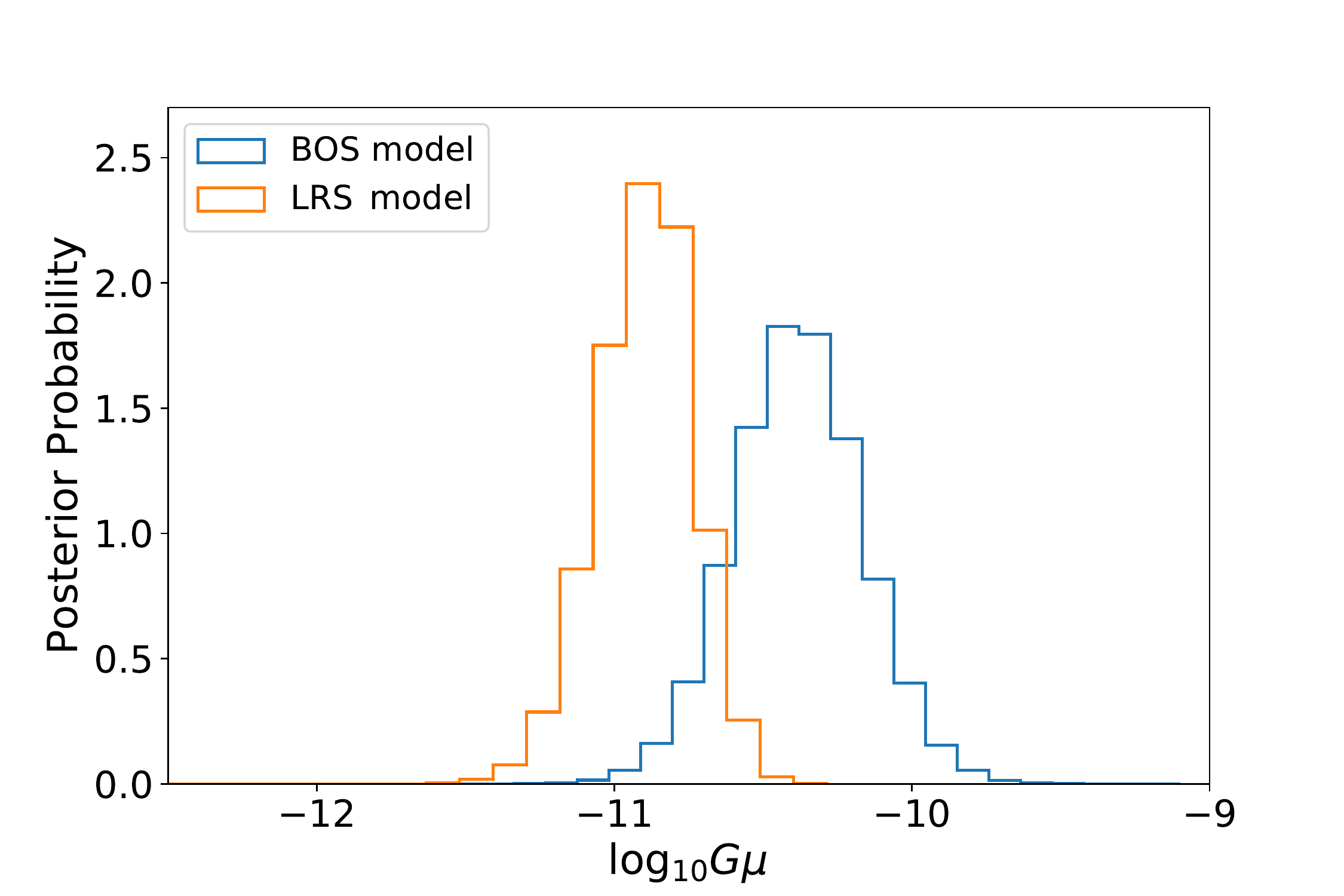}    
\caption{Posteriors distribution of $\log_{10}G\mu$ for hypothesis H2, for the BOS and LRS models.}
\label{CSH2}
\end{figure}

\begin{figure*}[!htp]
\centering
\includegraphics[width=0.48\textwidth]{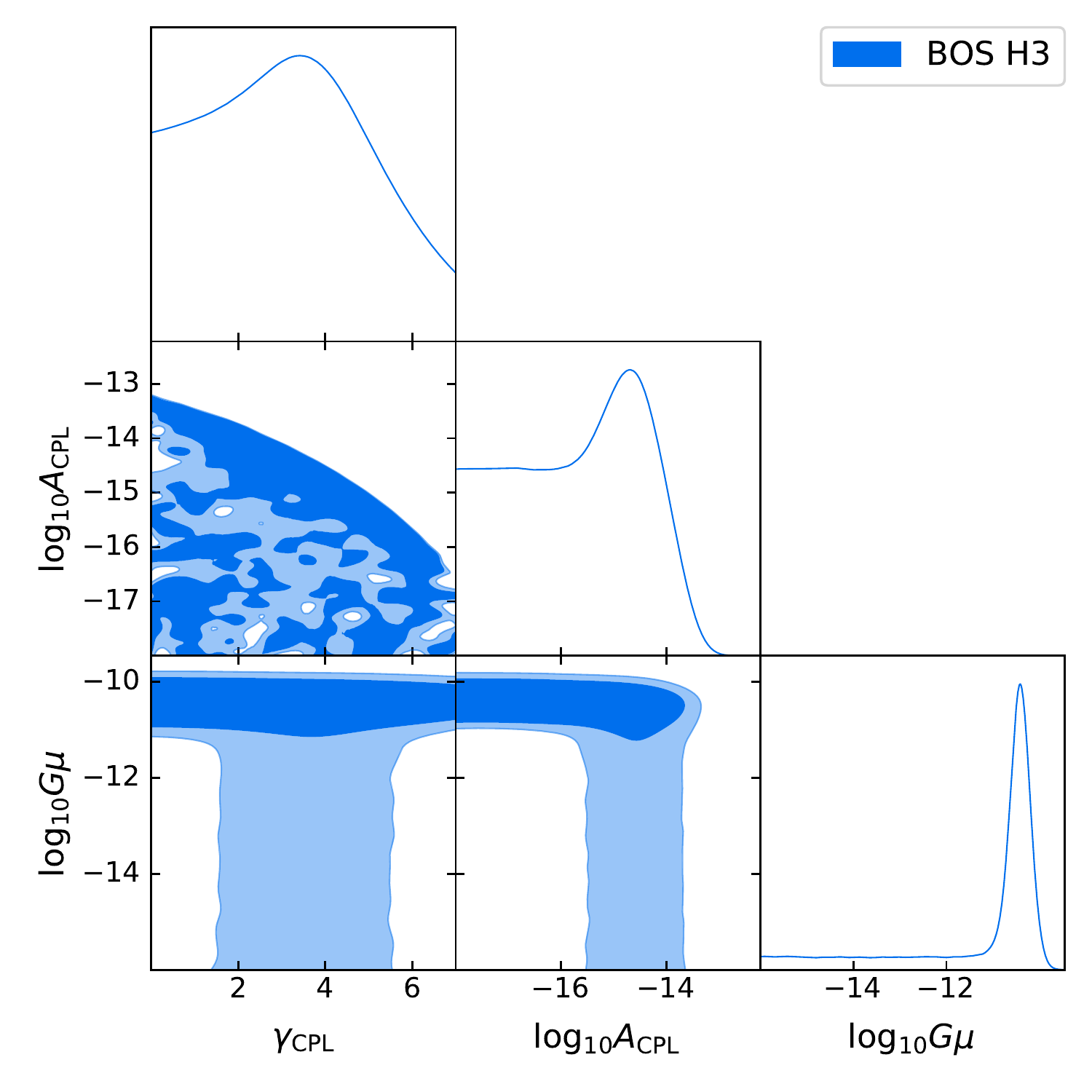}
\includegraphics[width=0.48\textwidth]{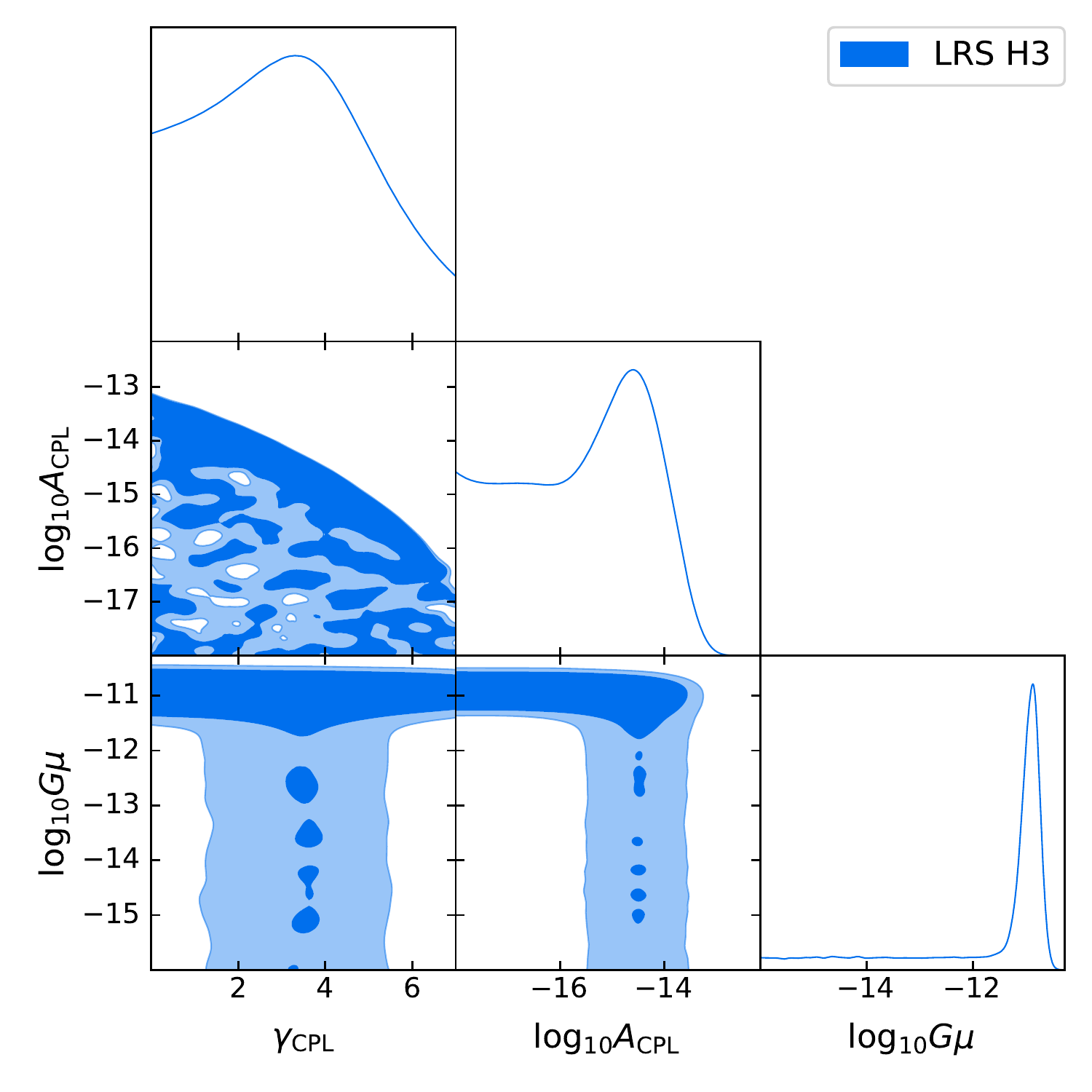}
\caption{Posterior distributions of $\log_{10}G\mu$ and the CPL parameters for hypothesis H3, 
for the BOS (left) and LRS (right) models.}
\label{H3BL}
\end{figure*}

Then we search for the signal of SGWB from the CS. Fig. \ref{CSH2} shows the one-dimensional
probability distributions of the CS parameter $G\mu$ without including the CPL process (H2).
This test reveals that the CS-induced SGWB can be used to explain the CPL excess in the data,
with similar BFs compared with the CPL assumption. 

Fig.~\ref{H3BL} gives the posterior distributions of the CPL parameters and the CS parameter
for hypothesis H3. In this case, the CS signal is not evident, and we can derive constraints
on the CS parameter. At the 95\% C.L., we find $\log_{10}G\mu<-10.02$ ($-10.64$) for the
BOS (LRS) model. Similar analysis but removing the auto-correlation of the pulsars' ToAs
(H4) is presented in the main text.

Finally we consider the possibility that the CPL has an astrophysical origin from the
supermassive binary black holes (SMBBH). The spectrum of the SMBBH SGWB signal is given by
\begin{equation}
S(f) =\frac{A_{\textrm{SMBBH}}^2}{12\pi^2}\left(\frac{f}{{\rm yr}^{-1}}\right)^{\gamma_{\textrm{SMBBH}}}~{\rm yr}^3,
\end{equation}
where the power-law index $\gamma_{\textrm{SMBBH}}$ is fixed to be 13/3. The prior of $\log_{10}A_{\textrm{SMBBH}}$ is assumed to be uniformly distributed in $[-18,-14]$. 
We test three model assumptions, with H5 being the SMBBH-only hypothesis and H6 (H7)
being the SMBBH + CS for the BOS (LRS) model. 

\begin{figure}[!htp]
\includegraphics[width=0.48\textwidth]{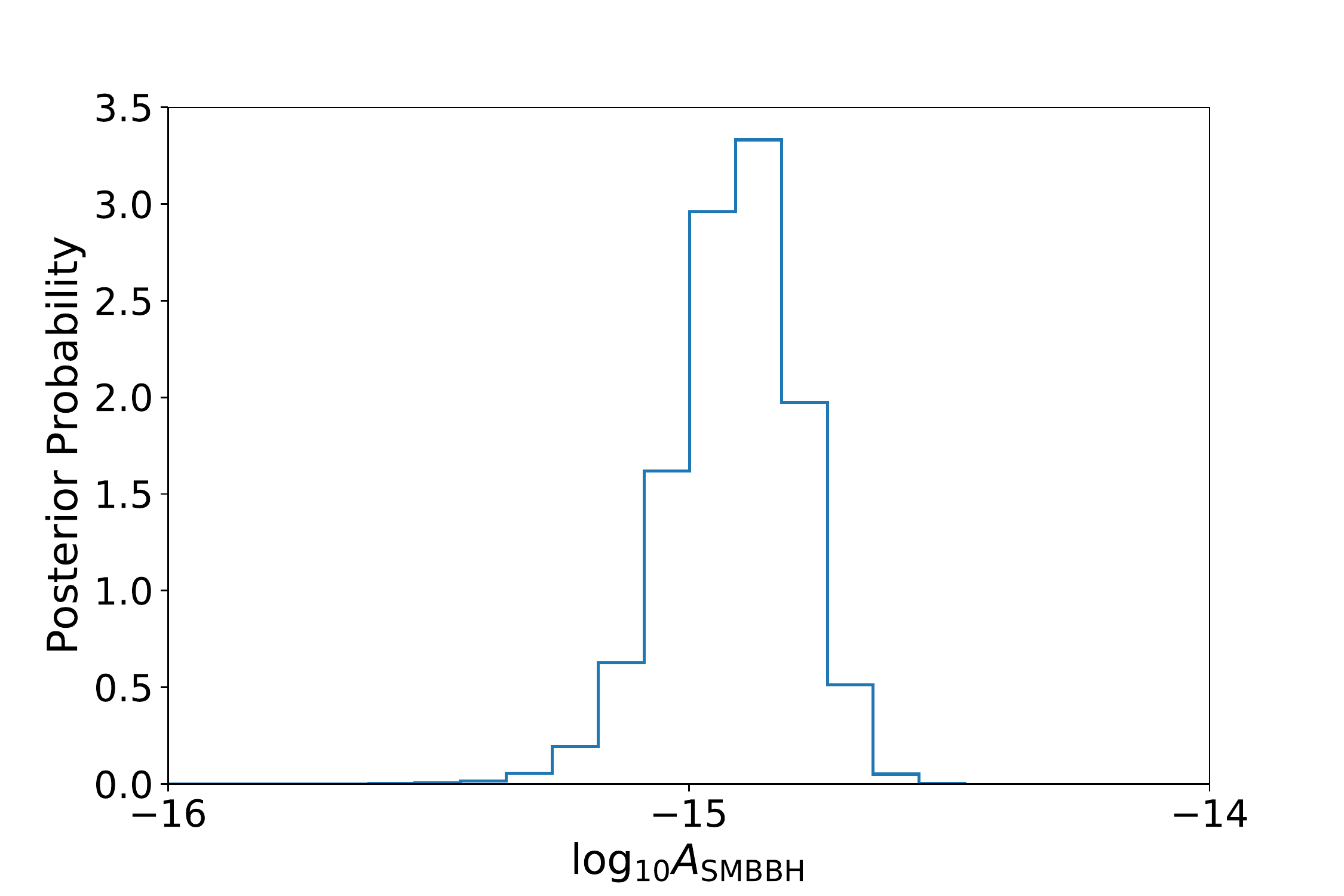}
\caption{Posterior distribution of $\log_{10}A_{\textrm{SMBBH}}$ for the SMBBH-only hypothesis.}
\label{onlySMBBH}
\end{figure}

\begin{figure*}[!htp]
    \centering
    \includegraphics[width=0.45\textwidth]{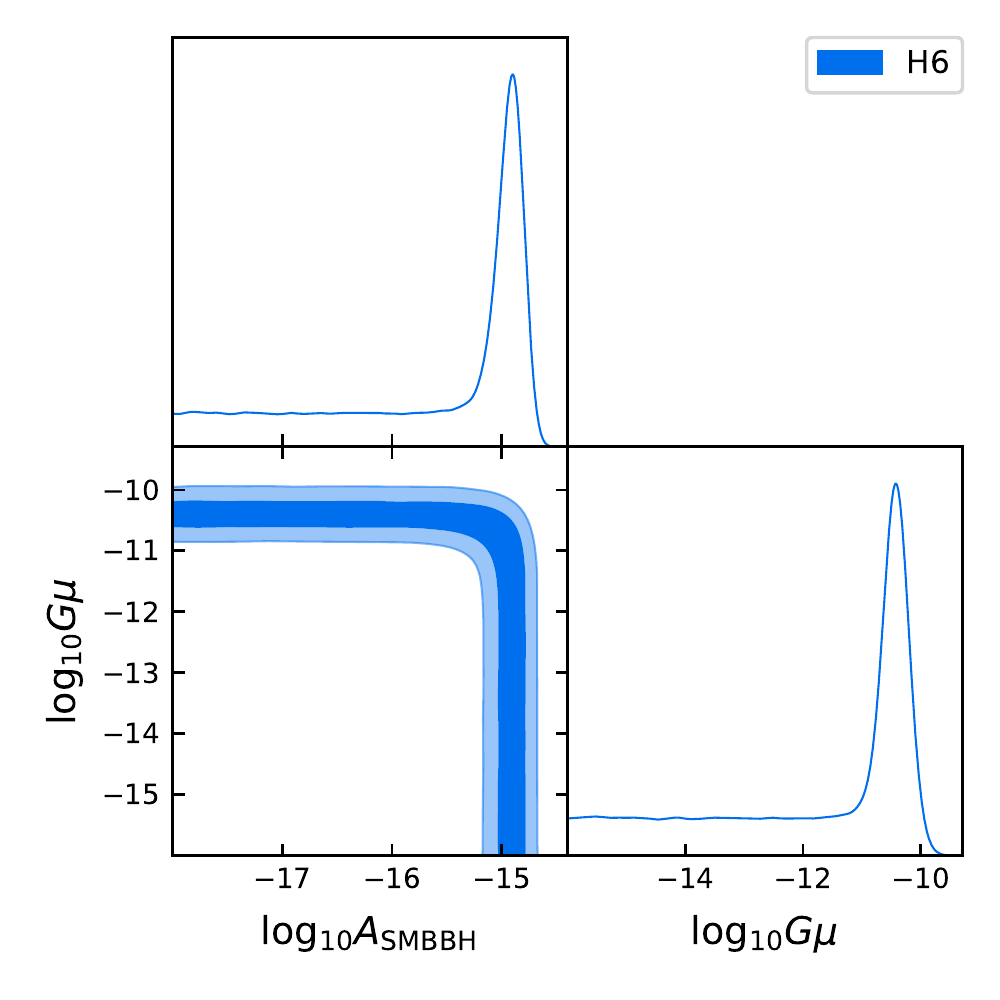}
    \includegraphics[width=0.45\textwidth]{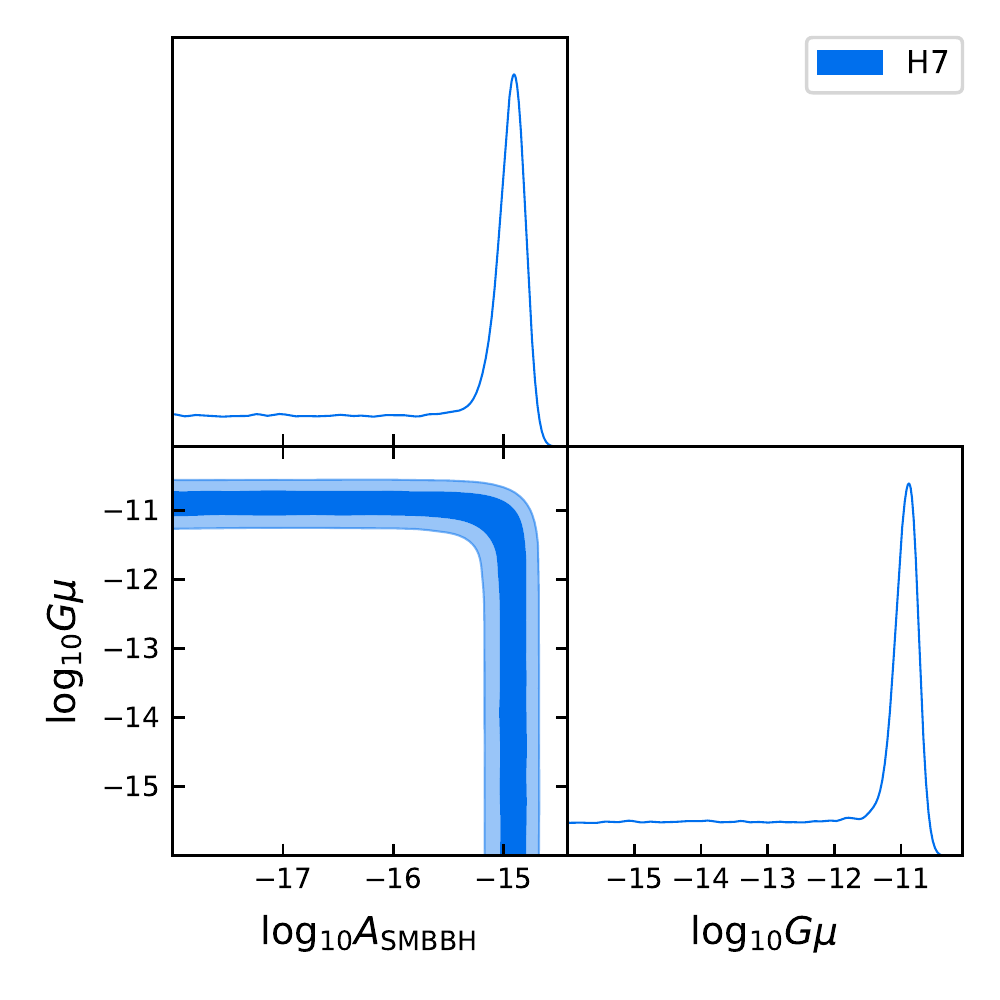}
    \caption{Posterior distributions of $\log_{10}G\mu$ and $\log_{10}A_{\textrm{SMBBH}}$ 
    for the BOS (left) and LRS (right) model.}
    \label{SMBBH}
\end{figure*}

The constraints on the model parameters of hypotheses H5 - H7 are shown in Fig.~\ref{onlySMBBH},
Fig.~\ref{SMBBH}, and Table \ref{SMBBHBYS}. The CPL process may also be explained by the SMBBH
signal, with $\log_{10} A_{\textrm{SMBBH}} = -14.89^{+0.10}_{-0.12}$. 
For hypotheses H6 and H7, we reach similar conclusion with H3 or H4 where the CPL is treated 
as a background but with slight changes of the 95\% C.L. upper limits of $\log_{10}G\mu$.

\begin{table*}[!ht]
\caption{Hypotheses, Bayes factors, and estimated model parameters of different model assumptions
for the case with the CPL substituted by the SMBBH.}
    \label{SMBBHBYS}
    \begin{tabular}{|l|c|c|c|c|c|c|}
    \hline 
    \multirow{2}{*}{Hypothesis} & 
    Pulsar &
    SMBBH &
    HD process  &
    \multirow{2}{*}{Bayes Factors}  & \multicolumn{2}{c|}{Parameter Estimation ($1\sigma$ interval)}\\
    \cline{6-7}
    & Noise & process  &CS spectrum & & $\quad\quad \log_{10}G\mu \quad\quad$ & $\log_{10}A_{\text{SMBBH}}$\\
    \hline
    H5:SMBBH & \checkmark & \checkmark &  & $ \ 10^{3.3} $ (/H0)  &  & $-14.89^{+0.10}_{-0.12}$ \\
    \hline
    H6:CS3 & \checkmark & \checkmark & \checkmark(full hd) & $ \ 1.21 $ (/H5) & $<-10.05$ ($95\%$ C.L.) & $-15.11^{+0.25}_{-1.88}$ \\
    \hline
    H7:CS4 & \checkmark & \checkmark & \checkmark(full hd) & $ \ 1.07 $ (/H5) & $<-10.66$ ($95\%$ C.L.) & $-15.09^{+0.24}_{-1.87}$ \\
    \hline
    \end{tabular}
\end{table*}

\end{document}